\documentclass[pdflatex,sn-mathphys]{sn-jnl}

\jyear{2021}%

\theoremstyle{thmstyleone}%

\usepackage{color}
\theoremstyle{thmstyletwo}%

\newcommand{\be}{\begin{equation}}
\newcommand{\ee}{\end{equation}}
\definecolor{pinegreen}{rgb}{0.0, 0.47, 0.44}

\theoremstyle{thmstylethree}%

\begin{document}

\title[]{Dominant Energy Condition and dissipative fluids in general 
relativity}


\author*[1]{\fnm{Valerio} \sur{Faraoni}}\email{vfaraoni@ubishops.ca}

\author[1,2]{\fnm{El Mokhtar Z. R.} \sur{ Mokkedem}}\email{emokkedem22@ubishops.ca}

\affil*[1]{\orgdiv{Department of Physics \& Astronomy}, 
\orgname{Bishop's University}, 
\orgaddress{\street{2600 College Street}, \city{Sherbrooke}, 
\postcode{J1M~1Z7}, \state{Qu\'ebec}, \country{Canada}}}

\affil[2]{\orgdiv{Unit\'e de Formation et de Recherche Physique, 
Ing\'enierie, Terre, Environnement, M\'ecanique}, 
\orgname{Universit\'e Grenoble Alpes}, 
\orgaddress{\street{126 Rue de la Physique}, 
\city{Saint-Martin-d'H\`eres}, \postcode{38400}, 
\country{France}}}

\abstract{Existing literature implements the Dominant Energy Condition 
for dissipative fluids in general relativity. It is pointed out that 
this condition fails to forbid superluminal flows, which is what it is 
ultimately supposed to do. Tilted perfect fluids, which formally have 
the stress-energy tensor of imperfect fluids, are discussed for 
comparison.}





\maketitle

\section{Introduction}
\label{sec:1}

The study of fluids in general relativity (GR) is well-developed and 
perfect 
fluids play a dominant role in this literature 
\cite{RezzollaZanotti,Andersson:2020phh}, but they do not  
describe situations in which dissipation becomes important such as, for 
example, the oscillation of neutron stars or  the generation of 
gravitational waves from compact objects \cite{Andersson:2002ch}. The 
most common model for dissipative relativistic fluids exhibits a purely 
spatial 
heat current density $q^a$, which is obviously non-causal and is the 
subject \cite{Maartens:1996vi,HisckockLindblom99,Andersson:2020phh}  of 
Eckart's first-order thermodynamics \cite{Eckart40}. This causality 
problem is cured in the Israel-Stewart and in other 
versions of second-order (causal) thermodynamics \cite{Muller67, 
Stewart77, 
IsraelStewart79a, IsraelStewart79b,Carter91,MullerRuggeri98} but, due to 
the inherent complication of these formalisms, the simplest 
non-causal model is still the most used in GR.

Relativistic forms of matter, including fluids, are supposed to satisfy 
energy conditions which forbid negative energy densities, superluminal 
mass and energy flows, and overly negative stresses. Indeed, without 
requiring any energy condition, one could write down any metric 
tensor $g_{ab}$ 
and, running the Einstein equations\footnote{We use the notation of 
Ref.~\cite{Waldbook}, in which the metric has signature ${-}{+}{+}{+}$ and 
units are used in which the speed of light $c$ and Newton's constant $G$ 
are unity.}
 \be 
R_{ab} - \frac{1}{2} \, g_{ab} R =8\pi T_{ab} 
\ee 
from left to right, compute the effective stress-energy 
tensor $T_{ab}$ that sources such a metric (here $R_{ab}$ denotes the 
Ricci tensor of the metric $g_{ab}$ and $R\equiv {R^c}_c$ is its trace). 
In general, this procedure produces senseless effective 
energy-momentum tensors $T_{ab}$ and completely unphysical solutions of 
the Einstein equations, hence the need to keep in check the physically 
admissible forms of matter described by  
$T_{ab}$ by imposing suitable energy conditions \cite{Waldbook,Carroll}.  
Moreover, energy conditions are crucial in the proofs of the black hole 
and 
cosmological singularity theorems \cite{HawkingEllis,Borde1987} and of the 
positivity of mass \cite{Bekenstein:1975wj}. The 
classical energy conditions are, however, doubted at least in the 
semiclassical 
context and are regarded as temporary requirements evolving as our 
knowledge of relativistic matter changes, usually on the scale of decades 
(\cite{Barcelo:2002bv}, see \cite{Martin-Moruno:2017exc} for a recent 
review).

Things become tricky when the energy conditions are contemplated for 
{\em dissipative} fluids. The literature on this subject is very 
limited: 
two articles \cite{Kolassis,Pimentel:2016jlm} address head-on the issue of 
energy conditions for imperfect fluids and they seem to have been quite 
influential.\footnote{At the time of writing, a 
cumulative citation count in Google 
Scholar returns 157 entries.} However, it is time to reconsider 
the energy 
conditions, and particularly the Dominant Energy Condition (DEC) for 
dissipative fluids. 
The main point here is that, while technically one can satisfy the energy 
conditions, the imperfect fluid model adopted unavoidably contains a 
purely spatial heat flux density vector. Specifically, this non-causal 
flow is not eliminated by imposing the usual DEC, which instead was 
originally supposed (and is naively believed) to 
eliminate superluminal flows completely. Thus, superficially 
it may seem reassuring that the DEC is satisfied but the usual 
dissipative fluid 
still, and unavoidably, contains non-causal heat propagation.

Another issue needs clarification. The DEC is universally taken to be 
meaningful for {\em perfect} fluids 
and understood as  
forbidding superluminal flows. Since perfect fluids are not 
dissipative, the 
problem 
mentioned above does not apply but, technically, the stress-energy tensor 
$T_{ab}$ of a perfect fluid that is tilted ({\em i.e.}, seen from the 
frame of an observer not comoving with this fluid) has the imperfect fluid 
form. This happens because the relative motion of the two 
frames generates a convective current in any frame in which this fluid is 
not at rest, and this has the same form as the heat flux of an imperfect 
fluid.  The DEC seems to take two different forms, the 
perfect fluid formulation in the comoving frame and the imperfect fluid 
DEC version in any non-comoving frame, leading to potential confusion, at 
least at first sight. We check explicitly that the two formulations 
of DEC, in fact, agree with each other for tilted perfect fluids.

\section{DEC and dissipative fluids}
\label{sec:2}

In GR, the imperfect fluid is characterized by the stress-energy tensor
\be
T_{ab}= \rho u_a u_b +P h_{ab} +\pi_{ab} + q_a u_b +q_b u_a \,,\label{Tab}
\ee
where $u^a $ is the fluid four-velocity satisfying $u_c u^c=-1$, 
\be
h_{ab} \equiv u_a u_b +g_{ab}
\ee
is the Riemannian metric on the 3-space orthogonal to $u^a$,  
$\rho$ is the energy density,
\be
P=\bar{P}+P_\mathrm{viscous}
\ee
is the isotropic pressure, consisting of  a non-viscous contribution 
$\bar{P}$ and of  a viscous one $P_\mathrm{viscous}$, $\pi_{ab}$ 
is the trace-free anisotropic stress-tensor, and $q^a$ is the heat flux 
density. $h_{ab}, \pi_{ab}, $ and $ q^a$ are purely spatial,
\be
h_{ab} u^a =h_{ab} u^b = 
\pi_{ab} u^a = \pi_{ab} u^b = q_a u^a=0 \,.
\ee

The Dominant Energy Condition (DEC) requires $
-T_{ab} v^b \leq 0$ for all timelike vectors $v^a$ and is supposed to 
forbid superluminal flows of mass-energy \cite{Carroll}. However, 
satisfying the DEC does not achieve this goal for imperfect 
fluids because the heat flux 
density vector, being purely spatial, will always describe a superluminal 
energy flow when it does not vanish.

Let us consider the special case in which the timelike vector $v^b$ in the 
formulation of the DEC is the four-velocity of the imperfect fluid 
itself, $v^a = u^a$. In this case, the corresponding energy flow is
\be
j_a \equiv -T_{ab} u^b =\rho u_a + q_a 
\ee
and consists of the material flow $\rho u_a$ plus the heat flow $q_a$. It 
is easy to see that
\be
j_a j^a = \left( \rho u_a +q_a \right) \left( \rho u^a +q^a \right) =  
-\rho^2 + q^2 \label{easy}
\ee
is non-positive, and $j^a$ is causal ({\em i.e.}, timelike or null),  if 
and only if $\rho^2 \geq q^2 \equiv q_c q^c$, which is part of the 
formulation of the DEC reported in Ref.~\cite{Pimentel:2016jlm} (to  
guarantee that  the vector $  S_a \equiv -T_{ab} v^b$   
is future-oriented). 
However, 
the vector $q^a$ remains purely spatial and non-causal. The DEC is 
enforced if, in addition \cite{Kolassis,Pimentel:2016jlm},
\be
\rho^2 \geq P_i^2 +q^2 +2 \left( \rho + 3 P \right) q \,, \quad  
i=1,2,3\,,\label{questa}
\ee
where the $P_i$ are the eigenvalues of the total stress tensor $\Pi_{ab} 
= P h_{ab} +\pi_{ab}$ and 
\be
P\equiv  \frac{1}{3} \, h^{ab} T_{ab}
\ee
is the (total) isotropic pressure. 
Even when~(\ref{questa}) is satisfied, the generic energy flux $-T_{ab} 
v^b $ 
contemplated in the DEC 
cannot 
forbid the heat flux vector 
\be
q_a = -{h_a}^c u^b T_{cb} \neq -u^b T_{ab} \equiv j_a    \,.
\ee 
It seems that a more effective formulation of the DEC would require that 
\begin{center} {\em both $ -T_{ab} v^b $ and $-{h_a}^c v^b T_{cb}$ are 
timelike or null for any timelike vector $v^a$,} \end{center} but this is 
tautological because $-{h_a}^c v^b T_{cb}$, being the spatial component of 
$j_a$, is necessarily non-causal if it is non-vanishing. The point is that 
$- T_{ab} v^b $ includes two {\em distinct} fluxes of very different 
nature, the convective flow of mass-energy $\rho u_a$ and the purely 
diffusive heat 
flux $q_a$. Forbidding the first from being superluminal does nothing to 
restrict the second, which remains superluminal. Thus, although the 
``total'' vector field $j_a = \rho u_a +q_a$ can be timelike, this is an 
artificial mathematical entity that does not catch the underlying physics  
and there remains a non-causal heat flux. In other words, the DEC 
restricts the magnitude $q$ of $q^c$ without changing its causal nature. 

Similar conclusions are reached by considering the Landau frame of the 
dissipative fluid.  The Landau (or energy, or Landau-Lifschitz) frame 
is the frame based on the direction of the total energy flux, hence the 
dissipation of energy does not appear explicitly in this frame, 
\be 
q_a^\mathrm{(L)} = -T_{cd} u^c_\mathrm{(L)} \, {h^\mathrm{(L)}_a}^d =0 
\ee 
where 
\be 
h^\mathrm{(L)}_{ab} \equiv g_{ab}+ u^\mathrm{(L)}_a u^\mathrm{(L)}_b \,, 
\ee 
but the energy flux is traded with 
a particle flux $N^a$. The four-velocity $u^a_\mathrm{(L)}$ defining the 
Landau frame, which is the direction of the total energy flux, is an 
eigenvector of the stress-energy tensor, 
\begin{equation}
    T_{ab} u_\mathrm{(L)}^b = -\rho_\mathrm{(L)} u^\mathrm{(L)}_a \,. 
\end{equation} 
In fact, in the Landau frame $q^a_\mathrm{(L)}=0$ and the stress-energy 
tensor of the imperfect fluid is decomposed (differently than in the 
Eckart frame) as 
\be 
T_{ab}= \rho_\mathrm{(L)} u^\mathrm{(L)}_a u^\mathrm{(L)}_b 
+ P^\mathrm{(L)} h^\mathrm{(L)}_{ab} +\pi^\mathrm{(L)}_{ab} \,. 
\ee 
Since the 
diffusive energy flux is already contained within the direction of 
$u^a_\mathrm{(L)}$, in this frame energy diffusion does not appear 
explicitly and the heat current is $q^\mathrm{(L)}_a = 0$. Then   $ 
u_\mathrm{(L)}^a $ is an eigenvector of $T_{ab}$: 
\begin{eqnarray} 
T_{ab} \, u_\mathrm{(L)}^b &=& \left( \rho_\mathrm{(L)} 
u^\mathrm{(L)}_a u^\mathrm{(L)}_b + P^\mathrm{(L)} h^\mathrm{(L)}_{ab} 
+\pi^\mathrm{(L)}_{ab} \right) u_\mathrm{(L)}^b \nonumber\\
&&\nonumber\\
&=& \rho_\mathrm{(L)} u^\mathrm{(L)}_a \left( u^\mathrm{(L)}_b 
u_\mathrm{(L)}^b \right)
+ h^{(L)}_{ab} u_{(L)}^b +\pi^\mathrm{(L)}_{ab} u_\mathrm{(L)}^b 
  \nonumber\\
&&\nonumber\\
&=& -\rho_\mathrm{(L)} u^\mathrm{(L)}_a \,.
\end{eqnarray}
When the strong energy condition holds, the timelike four-velocity 
eigenvector $u_\mathrm{(L)}^a$ of the imperfect fluid stress-energy tensor 
is unique \cite{EllisMaartensMcCall} (the uniqueness of the 
Landau frame has been derived also from the relativistic Boltzmann 
equation using renormalization group techniques \cite{Tsumura:2012ss}). 
The particle flux density is 
\begin{equation}
    N^a = n_\mathrm{(L)} \, u^a_\mathrm{(L)} + V^a_\mathrm{(L)} \,, 
\end{equation} 
where  $V^a_\mathrm{(L)}$ is the particle diffusion current density.

The requirement that the formal mass-energy current density 
$-T_{ab} u^b_\mathrm{(L)} $ be timelike (respectively, null) is 
achieved for timelike (null)  four-velocity $u^a_\mathrm{(L)}$ since
\be 
 \left( -T_{ab} u^b_\mathrm{(L)}\right) \left( -{T^a}_c 
u^c_\mathrm{(L)}\right) = \rho^2_\mathrm{(L)} \left( u^\mathrm{(L)}_a 
u_\mathrm{(L)}^a \right) \leq 0 
\ee 
implies $u^\mathrm{(L)}_a u_\mathrm{(L)}^a \leq 0$. However, the 
non-causal nature of  the heat flux is now traded with the non-causal 
nature of the particle flux since $V^a_\mathrm{(L)}$ is spacelike. The 
relation between  quantities in the Eckart and the Landau frames is 
\begin{equation}
    u^\mathrm{(E)}_a = \frac{N_a}{n_\mathrm{(E)} } = 
\frac{n_\mathrm{(L)} u_a^\mathrm{(L)} + 
V_a^\mathrm{(L)}}{n_\mathrm{(E)}} \,.
\end{equation}
Let $V_a^\mathrm{(L)} V^a_\mathrm{(L)} \equiv \alpha^2$, then
\begin{eqnarray}
n_\mathrm{(E)} &=& \sqrt{-N^c N_c} = \sqrt{n^2_\mathrm{(L)} - \alpha^2} 
= 
n_\mathrm{(L)} \sqrt{1-\left(\frac{\alpha}{n_\mathrm{(L)} }\right)^2} 
\nonumber\\
&&\nonumber\\
&=& 
n_\mathrm{(L)}\sqrt{1-v^2}
\end{eqnarray}
where $v \equiv \alpha/ n_\mathrm{(L)}$ is the relative velocity 
between the 
Eckart and Landau frames (a purely spatial 
vector) \cite{EllisMaartensMcCall}. One can now have $u^a_\mathrm{(L)}$ 
and $N^a =n_\mathrm{(E)} \, u_\mathrm{(E)}^a -V^a_\mathrm{(L)}$  
timelike, but 
$V^a_\mathrm{(L)}$ is spacelike.

The new definition of DEC requiring both $-T_{ab} v^b$ and its spatial 
projection 
to be causal for any timelike $v^b$ would simply prohibit one {\em a 
priori} from 
considering imperfect fluids with spacelike heat flux density $q^a$. It 
seems rather pointless, therefore, to impose the usual DEC on imperfect 
fluids of the form~(\ref{Tab}) with $q^c q_c >0$.

An alternative stress-energy tensor  used in models of anisotropic 
spherical stars is given, in spherical coordinates, by  
\cite{Setiawan:2019ojj} 
\be
T_{ab}^{(2)}= 
\rho u_a u_b +P_t h_{ab} +\sigma \, q_a q_b \,,
\ee
where 
\be
\sigma = P_t -P_r
\ee
is the difference between the tangential and radial pressures $P_t$ and 
$P_r$  
and the purely spatial vector $q^a$ is interpreted as a a radial velocity 
\cite{Setiawan:2019ojj}. This interpretation is, however, questionable 
because the four-velocity of the fluid is $u^a$ and this fluid cannot have 
simultaneously two velocities, in addition to the fact that one would  
have a spacelike radial four-velocity. In this case, 
$-T_{ab} u^b = \rho u_a$ is a timelike vector. This fluid is 
anisotropic but apparently non-dissipative, so it really fails to address 
our problem 
of the DEC's relation with dissipative fluids.

\section{Tilted perfect fluid}
\label{sec:3}

There is another occurrence of imperfect fluid which, in reality, is   
only a perfect (non-dissipative) fluid in disguise.  A perfect fluid with 
stress-energy tensor
\be
T_{ab}=\rho^* \, u^*_{a} \, u^*_{b} +P^* h^*_{ab} \,,\label{perfectfluid}
\ee
when seen from a non-comoving frame, {\em i.e.}, from a frame based on 
observers with a (timelike) four-velocity $u^a$ different from the fluid 
four-velocity 
$u^*_a$,  or ``tilted'', appears as a dissipative fluid 
\cite{EllisMaartensMcCall,Maartens:1998xg}. The comoving frame is the 
unique  frame in which the perfect fluid stress-energy tensor assumes 
the 
perfect fluid form~(\ref{perfectfluid}) 
\cite{EllisMaartensMcCall,Maartens:1998xg}. 
In the frame of  a different observer with timelike four-velocity $u^a$ 
related to $u^{*a} $ by \cite{EllisMaartensMcCall,Maartens:1998xg}
\begin{eqnarray}
u^{* a} &=& \gamma \left( u^a+v^a \right) \,,\label{frame1}\\
&&\nonumber\\
\gamma &=& \frac{1}{\sqrt{ 1-v^2}}=-u^*_c u^c  \,,\label{frame2}\\
&&\nonumber\\
v^2 &\equiv & v^c v_c \,, \quad \quad v^c 
u_c = 0 \,, \quad \quad 0 \leq v^2<1 \,,\label{frame3}
\end{eqnarray}
this perfect fluid (now ``tilted'') will appear as a dissipative fluid 
with the different stress-energy tensor decomposition 
\cite{EllisMaartensMcCall,Maartens:1998xg,Clarkson:2003ts,Clarkson:2010uz} 
\be
T_{ab}=\rho  \, u_a \, u_b +P h_{ab} 
+q_a u_b +q_b u_a +\pi_{ab} \,,
\ee
where $h_{ab}= g_{ab}+ u_a u_b$ as usual and 
\cite{EllisMaartensMcCall,Maartens:1998xg,Clarkson:2003ts,Clarkson:2010uz}  
\begin{eqnarray}
\rho &=& \rho^*  + \gamma^2 \,  v^2 \left( \rho^* + P^* \right)
= \gamma^2  \left( \rho^* + v^2 P^* \right)\,,\label{sft1}\\
&&\nonumber\\
&&\nonumber\\
P &=& P^*  +\frac{\gamma^2 \, v^2}{3}  \left( \rho^* + P^* \right)  
\,,\label{sft2}\\
&&\nonumber\\
&&\nonumber\\
q^a &=&  \left( 1+  \gamma^2 \, v^2 \right) \left( \rho^* + P^* 
\right) v^a \nonumber\\
&&\nonumber\\
&=& \gamma^2  \left( \rho^* + P^* \right) v^a     \,,\label{sft3}\\
&&\nonumber\\
&&\nonumber\\
\pi^{ab} &=&  \gamma^2 \left( \rho^* + P^* \right) 
\left( v^a \, v^b -\frac{v^2}{3} \, h^{ab} \right) 
\,.\label{sft4}
\end{eqnarray}
That is, the {\em same} stress-energy $T_{ab}$ admits infinitely many 
{\em decompositions}: one based on the fluid four-velocity $u^{*a} $ and 
others based on different timelike vectors $u^a$.

Equation~(\ref{sft3}) lends itself to a natural physical interpretation. 
The heat current density $q^a$ is the energy crossing the unit of normal 
area per unit time. In Newtonian physics this flux density would be 
$\rho^* \vec{v}$: in GR also the pressure gravitates, hence 
$\rho^*$ is replaced by $\left( \rho^* +P^* \right)$. Then, the 
gravitating energy density is corrected by two Lorentz factors, one 
because of Lorentz contraction in the direction of motion (one of the 
three 
directions forming spatial volume, while the two directions perpendicular 
to the 
motion are not Lorentz-contracted). Then, this energy is blueshifted 
because  of time dilation associated with the motion, which contributes 
the second 
Lorentz factor $\gamma$. It is clear that now the spatial vector $q^a$ 
arises solely due to the relative motion between the two frames, {\em 
i.e.}, 
to the (spatial) vector $ v^a$. In this context it would be incorrect  to 
interpret this purely convective  flux as due to heat conduction, 
which should instead be given by Eckart's generalization of Fourier's law 
\cite{Eckart40}
\be
q_a= -K  h_{ab}\left( \nabla^b T + T \, \dot{u}^b \right) 
\,,
\ee
where $K$ is the thermal conductivity and $\dot{u}^a \equiv u^c \nabla_c 
u^a$ is the fluid's four-acceleration. In other words, the non-dissipative 
nature of the fluid~(\ref{perfectfluid}) remains in other (non-comoving)  
frames in spite of the form of the stress-energy tensor in these frames, 
which technically has the form~(\ref{Tab}) of a dissipative fluid  but 
ultimately is non-dissipative. To corroborate this argument, one notes 
that in Eckart's theory \cite{Eckart40} (as well as in the ordinary 
description of three-dimensional non-relativistic Newtonian fluids), the 
viscous pressure is assumed to be linear in the velocity gradient,
\be
P_\mathrm{viscous}=-\zeta \Theta = -\zeta \nabla_c u^c 
\,,\label{bulkviscosity} 
\ee
where $\zeta$ is the bulk viscosity coefficient and $\Theta =\nabla_c u^c 
$ is the expansion scalar. Since 
\be
\Theta= \nabla_c u^c = \frac{\Theta^*}{\gamma} - 2\gamma \,v^a  u^{*c} 
\nabla_c v_a - \nabla_c v^c \,,
\ee
there is no natural way to relate the viscous pressure $ 
P_\mathrm{viscous}$ in the non-comoving frame with the expansion scalar 
$\Theta$ to satisfy the constitutive relation~(\ref{bulkviscosity}) for 
Newtonian 
dissipative fluids in this frame, another pointer to the fact that there 
is no real heat flux. Similarly, there appears to be no way to satisfy the 
other constitutive relation of Eckart's thermodynamics \cite{Eckart40} $ 
\pi_{ab}=-2\eta \sigma_{ab}$ ({\em i.e.}, the anisotropic stresses being 
proportional to the shear tensor $\sigma_{ab}$ through a viscosity 
coefficient $\eta$) with a purely convective energy flux. 

Imposing the DEC on the perfect fluid~(\ref{perfectfluid}) guarantees that 
DEC is satified in {\em any} frame in which this fluid is tilted and in 
which, accordingly, the formulation of DEC should be different from that 
for perfect fluids \cite{Kolassis,Pimentel:2016jlm}. In fact, 
assume that DEC is satisfied in the comoving frame of the perfect fluid 
with four-velocity $u^{*a}$. Any other given timelike vector $u^a$ can 
be related to $u^{*a}$ by Eqs.~(\ref{frame1})-(\ref{frame3}). Then 
consider the flux $-T_{ab} u^b$ by decomposing the stress-energy tensor 
$T_{ab}$ with respect to this vector $u^a$. The DEC expressed using the 
arbitrary timelike vector $ u^a $ says that $j_a \equiv -T_{ab} u^b 
=\rho u_a +q_a $ is causal and future-oriented, or $q^2-\rho^2 \leq 0$ 
according to 
Eq.~(\ref{easy}).

Using the transformation properties (\ref{sft1})-(\ref{sft3}), one obtains
\be
q^2 \equiv q_a q^q = \gamma^4 \left( \rho^* +P^* \right) v^2 
\ee
and 
\begin{eqnarray}
q^2 -\rho^2 &=& \gamma^4 \left( \rho^* +P^* \right) v^2
-( \rho^*)^2 - \gamma^4 v^4 \left( \rho^* +P^* \right)^2 \nonumber\\
&&\nonumber\\
&\, & 
-2\gamma^2 v^2 \rho^* \left( \rho^* +P^* \right) \nonumber\\
&&\nonumber\\
&=& \gamma^4 v^4 \left( \rho^* +P^* \right)^2\left(1-v^2 \right) 
-( \rho^*)^2 \nonumber\\
&&\nonumber\\
&\, &  -2\gamma^2 v^2 \rho^* \left( \rho^* +P^* \right) \nonumber\\
&&\nonumber\\
&=& - ( \rho^*)^2 + \gamma^2 v^2 \left[ (P^*)^2 -(\rho^*)^2    \right] 
\nonumber\\
&&\nonumber\\
& \leq &  - ( \rho^*)^2<0
\end{eqnarray} 
assuming $P^* \geq - \rho^*$ for the perfect fluid in the comoving frame. 
In this case, 
there is no real heat flux but only a heat-like purely convective energy 
flux and the usual DEC requirement suffices to 
forbid non-causal flows of mass-energy.

\section{Conclusions}
\label{sec:4}

As a conclusion, it seems rather pointless to impose the DEC on 
first-order dissipative fluids \`a la Eckart \cite{Eckart40}, in the 
sense that one can compute it for dissipative 
matter and formally satisfy it, but it loses its conventional physical 
meaning learned by studying non-dissipative matter. It still makes 
sense, though,  
to impose the DEC on fluids that {\em appear} dissipative because they are 
tilted perfect fluids observed in non-comoving frames. In this case, the 
DEC does a good job at forbidding non-causal flows {\em of material} 
and there is no diffusive (as opposed to convective) energy flow. This 
situation does not correspond to a true dissipative fluid because the 
energy flux is purely convective and not thermal, and this case is already 
covered by 
the DEC studied for perfect fluids. Thermal heat flow, however, has a 
different physical nature and eludes the usual DEC. Once a purely spatial 
heat flux is included in the stress-energy tensor $T_{ab}$, it will haunt 
the dissipative fluid forever, except in very special situations in which 
the heat flux is forced to vanish due, for example, to symmetries (as in 
Friedmann-Lema\^itre-Robertson-Walker cosmology where $q^a \neq 0$ would 
violate spatial isotropy). This situation highlights that the DEC 
restricts flows of material but not all energy fluxes.


\bmhead{Acknowledgments}

This work is supported, in part, by the Natural Sciences \& Engineering 
Research Council of Canada (grant 2016-03803 to V.F.).

\section*{Declarations}

The authors declare no conflict of interest. There are no data 
associated with this work because of its theoretical and mathematical 
nature. All authors contributed to the study conception and design.






\end{document}